\documentclass[preprint,superscriptaddress,16pt]{revtex4-1}

\usepackage{graphicx,natbib}
\usepackage{dcolumn}
\usepackage{bm}
\usepackage{amsmath,graphics,amssymb,psfrag}
\usepackage[dvipdfm,colorlinks=true,linkcolor=blue,urlcolor=blue,citecolor=blue]{hyperref}
\usepackage[left]{lineno}
\begin{document}

\title{The suppressed radiative recombination rate in a quantum photocell with three electron donors}

\author{Jing-Yi Chen }
\affiliation{Department of Physics, Faculty of Science, Kunming University of Science and Technology, Kunming, 650500, PR China}

\author{Shun-Cai Zhao}
\email[Corresponding author: ]{zhaosc@kmust.edu.cn }
\affiliation{Department of Physics, Faculty of Science, Kunming University of Science and Technology, Kunming, 650500, PR China}



\begin{abstract}
The radiative recombination of electron-hole pairs represents a great challenge to the photon-to-charge efficiency in the photocell. In this paper, we investigate how to suppress radiative recombination rate (RRR) in a proposed quantum photocell with three dipole-dipole coupled and uncoupled electron donors. The results showed that the RRR could be suppressed in this photocell with three uncoupled electron donors but be enhanced with three dipole-dipole coupled electron donors by the ambient circumstance temperatures, and the increasing energy gap in the donors, the decreasing gap between the donors and acceptor inhabited the RRR with three dipole-dipole both coupled and uncoupled electron donors. When the photocell was manipulated by the electrostatic dipole-dipole coupling strength \(J\) at room temperature, the RRR was suppressed to a smaller minimum by the gap between the donors and acceptor than those by different gaps in the donors. These suppressed strategies for RRR point out some significant ways to increase the photon-to-charge efficiency and deserve the further experimental verification.
\begin{description}
\item[PACS numbers]42.50.Gy; 42.50.Ct; 32.80.Qk
\item[Keywords]Quantum photocell; radiative recombination rate (RRR); photon-to-charge efficiency
\end{description}
\end{abstract}
\maketitle
\section{Introduction}

The photon-to-charge conversion efficiency\cite{1} is an important aspect of photocell. However, the radiative upward transition and its reversal, the radiative downward transition coexist simultaneously, which produced the detailed balance limit\cite{2} in 1961 by Shockley and Queisser. And the radiative recombination has been considered as the fundamental limit\cite{2} on the conversion efficiency and was accepted widely in the artificial light-harvesting systems. But beyond that, other energy loss processes, such as surface reflection, internal resistance, thermalization losses, unabsorbed photons with energy less than band gap\cite{3} still exit in the photocell. And many of these unessential energy loss processes can be minimized by appropriately designed structures\cite{4,5,6,7,8}, such as multi-junction\cite{9,10,11} and intermediate band photocells \cite{12,13,14,15,16}, etc..

Recently, theoretical and experimental studies\cite{17,18,19} have demonstrated that the quantum coherence can alter the conditions of the detailed balance between the absorption and radiative recombination, and thereby the suppressed radiative recombination enhances the conversion efficiency in quantum photocells\cite{20,21,22,23}. One of the possible ways suggested by Scully\cite{17} is to cancel the emission processes via the quantum coherence induced by an external source\cite{18}. Consequently, the quantum coherence of the delocalized donor states alters the conditions for the thermodynamic detailed balance, and then brings out the enhanced efficiency in the photocell\cite{20,21,22,23}.

Considering a higher conversion efficiency achieved in the triple-junction photocells\cite{24,25}, in the following, we focus on RRR in a proposed quantum photocell with three p-n junctions simulated by three electron donors and discuss the approach to inhabit the RRR, which may bring out the enhanced photon-to-charge efficiency. The results indicate some positive significance and some encouraging trends, which may attract the further experimental investigations.

The work is organized as follows: in section 2, we describe the quantum photocell model with three electron donors. And we present the results and the corresponding discussions regarding the RRR dependent some parameters and possible experimental realization in section 3. A concise summary is given in the final section.

\section{Quantum photocell model with three electron donors}

\begin{figure}
\centerline{\includegraphics[width=0.45\columnwidth]{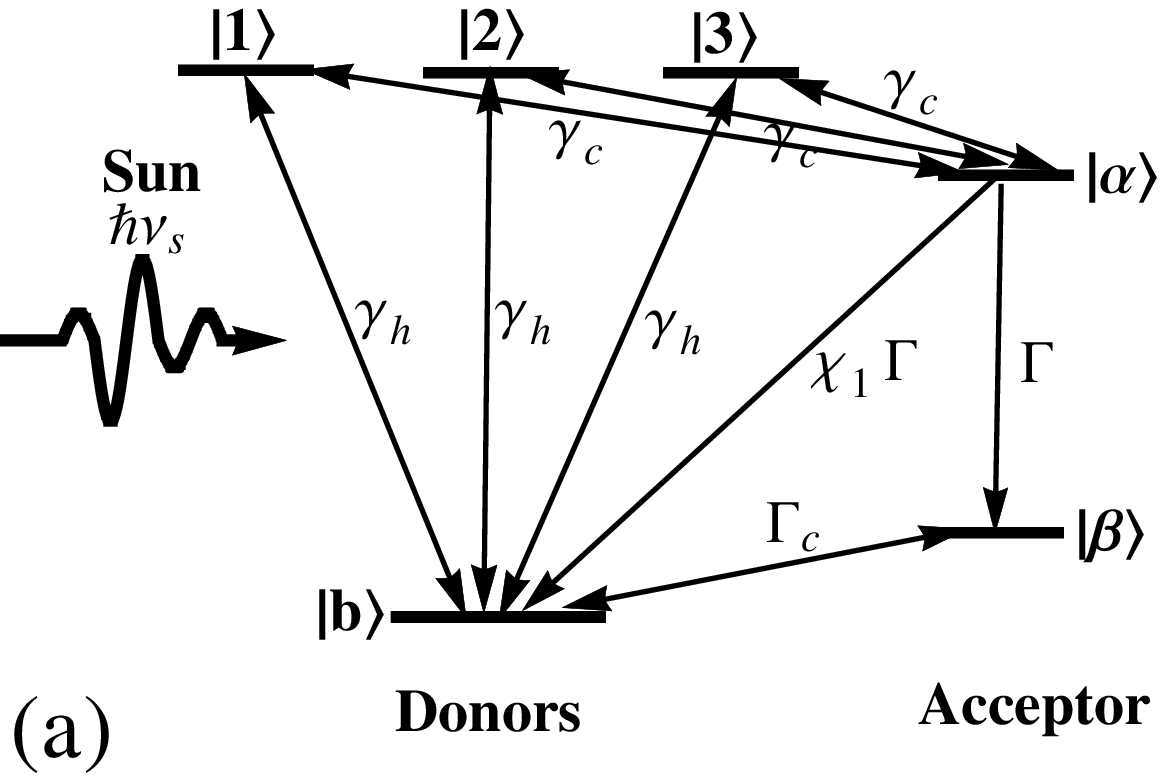}\includegraphics[width=0.45\columnwidth]{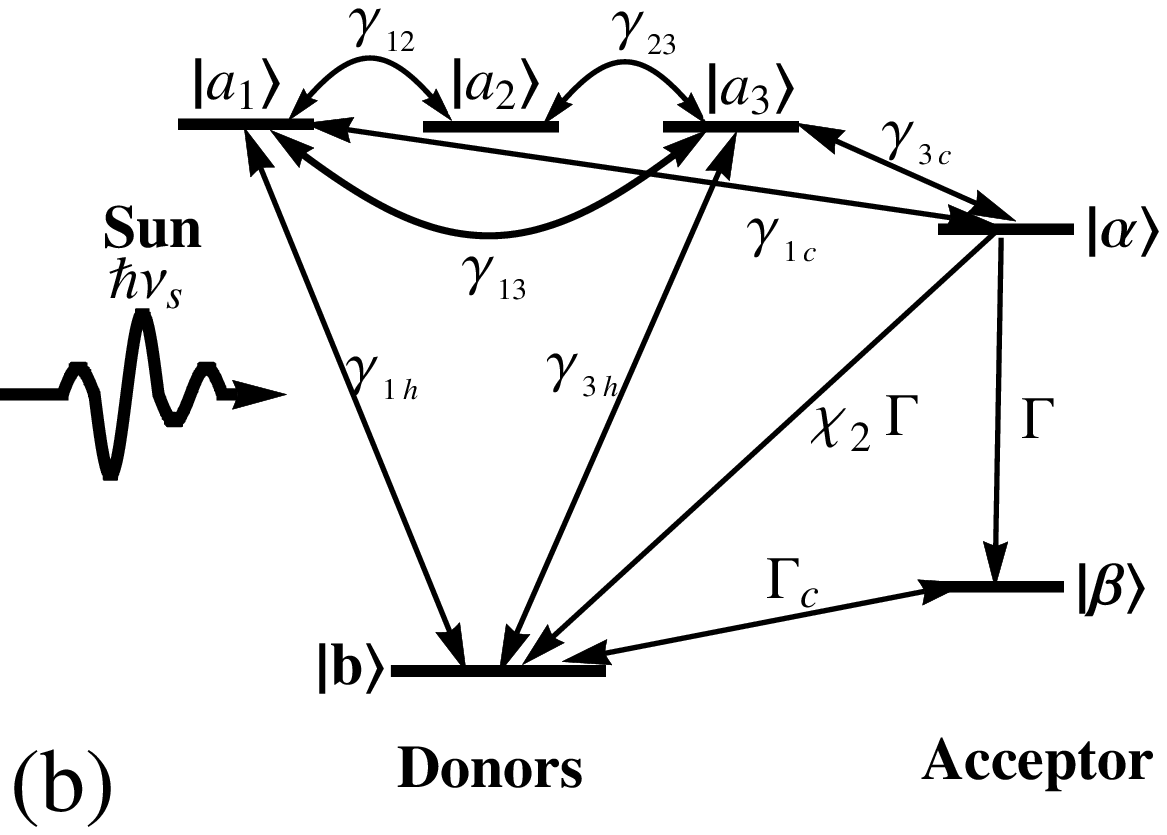}}
\caption{Schematic diagram: quantum photocell models with the acceptor and (a) three uncoupled donors, (b) three dipole-dipole coupled donors. Solar radiation drives electron transport between the valence band (VB) state \(|b\rangle\) and the conduction band (CB) state \(|i\rangle_{(i=1,2,3)}\) in Fig.(a). Transitions between levels \(|i\rangle_{(i=1,2,3)}\) \(\leftrightarrow\) \(|\alpha\rangle\), \(|\beta\rangle\) \(\leftrightarrow\) \(|b\rangle\) are driven by ambient thermal phonons. Levels \(|\alpha\rangle\) and \(|\beta\rangle\) are connected to a load. The three degenerate excited levels in Fig.1(a) split into Fig.1(b) because of the couplings between three donors, and the dark level \(|a_{2}\rangle\) is optically forbidden and has no electron transfer path to the acceptor \(|\alpha\rangle\).}
\label{f1}
\end{figure}

Proceeding with the analysis, we consider a quantum photocell model with the conduction band (CB) states \(|i\rangle_{(i=1,2,3)}\) and the valence band (VB) state  \(|b\rangle\) [depicted in Fig.1(a)] as the donors. And level \(|\alpha\rangle\) and level \(|\beta\rangle\) connecting to a load are assumed the acceptor molecule. The excitation of a molecule is simply modeled as a two-level system with the excited state \(|i\rangle_{(i=1,2,3)}\) and the ground state \(|b\rangle\). Then the excited electrons driven by solar radiation can be transferred to the acceptor molecule, the conduction reservoir state \(|\alpha\rangle\), with any excess energy radiated as a phonon into the ambient thermal phonons reservoirs. The excited electron is then assumed to be used to perform work, leaving the conduction reservoir state \(|\alpha\rangle\) decaying to the sub-stable state \(|\beta\rangle\) at a rate \(\Gamma\). The recombination between the acceptor and the donor is modeled as \(\chi_{i}\Gamma\) (i=1,2) in Fig.1, where \(\chi_{i} \) is the RRR, a dimensionless fraction. The recombination process brings the system back into the VB state \(|b\rangle\) without producing a work current, which could be a significant source of inefficiency. Finally, the state \(|\beta\rangle\) decays back to the VB state \(|b\rangle\) at a rate \(\Gamma_{c}\) and the cycle terminates. In Fig.1(a), the three donors are assumed to be identical and degenerate, and their three  uncoupled excited states \(|i\rangle_{(i=1,2,3)}\) have the same excitation levels \(E_{1}\)=\(E_{2}\)=\(E_{3}\)=E, and their transition dipole moments are aligned in the same direction, i.e., \(\vec{\mu}_{i}\)=e\(\langle i |\vec{r}|b\rangle_{(i=1,2,3)}\)=\(\vec{\mu}\), where \(\vec{r}=\vec{r}_{b}-\vec{r}_{i}\), and \(\vec{\mu}_{i}\) is located at \(\vec{r}_{i}\). The dipole-dipole interaction only exists in the nearest neighbors and the dipole-dipole couplings are denoted by J=\(\frac{1}{4\pi\epsilon\epsilon_{0}}[\frac{\vec{\mu}_{i}\cdot\vec{\mu}_{j}}{r^{3}}-\frac{3(\vec{\mu}_{i}\cdot \vec{r})(\vec{\mu}_{j}\vec{r})}{r^{5}}]\) between  \(|a_{1}\rangle \) and \(|a_{2}\rangle \), and \(|a_{2}\rangle \) and \(|a_{3}\rangle \) in Fig.1(b), but there is no coupling between \(|a_{1}\rangle \) and \(|a_{3}\rangle \). The strength of the dipole-dipole coupling J is much weaker than the excitation energy \(E- E_{b}=\hbar\omega\). The Hamiltonian of the three coupling donors can be written as

\begin{eqnarray}
&\hat{H}=&\sum^{3}_{i} \hbar\omega {\hat{\sigma}}^{\dag}_{i} {\hat{\sigma}}_{i} + J(\hat{{\sigma}}^{-}_{1}{\hat{\sigma}}^{\dag}_{2}+{\sigma}^{-}_{2}{\sigma}^{\dag}_{3}+H.c.),
\end{eqnarray}

\noindent where H.c. means Hermitian conjugation, \( \hat{\sigma}^{\dag}_{i}\) and  \( \hat{\sigma}^{-}_{i}\) are the Pauli raising and lowering operators, respectively. The three single-excitation states of the above Hamiltonian are \(|a_{1}\rangle\) =\(\frac{1}{2}\)(\(|1\rangle\)+\(\sqrt{2}|2\rangle\)+\(|3\rangle\)), \(|a_{2}\rangle\) =\(\frac{1}{\sqrt{2}}\)(\(|1\rangle\)-\(|3\rangle\))
,  \(|a_{3}\rangle\) =\(\frac{1}{2}\)(\(|1\rangle\)-\(\sqrt{2}|2\rangle\)+\(|3\rangle\)), and their eigenvalues are obtained as
\(E_{a_{1}}\)=\(E\)+\(\sqrt{2}\)J, \(E_{a_{2}}\)=\(E\), \(E_{a_{3}}\)=\(E\)-\(\sqrt{2}\)J. The dynamics behavior of the donors-acceptor system can describe via the master equations for the three uncoupled donors case in Eqn(2)(shown in Fig.1(a)) and the three dipole-dipole coupled donors case in Eqn(3)(shown in Fig.1(b)) as follows, respectively.
\begin{eqnarray*}
&\dot{\rho_{11}}=&\gamma_{h}[n_{h}\rho_{bb}\!-\!(1+n_{h})\rho_{11}]\!+\!\gamma_{c}[n_{c}\rho_{\alpha\alpha}\!-\!(1+n_{c})\rho_{11}],\\
&\dot{\rho_{22}}=&\gamma_{h}[n_{h}\rho_{bb}\!-\!(1+n_{h})\rho_{22}]\!+\!\gamma_{c}[n_{c}\rho_{\alpha\alpha}\!-\!(1+n_{c})\rho_{22}], \\
&\dot{\rho_{33}}=&\gamma_{h}[n_{h}\rho_{bb}\!-\!(1+n_{h})\rho_{33}]\!+\!\gamma_{c}[n_{c}\rho_{\alpha\alpha}\!-\!(1+n_{c})\rho_{33}],  ~~~~~~~~~~~~~~(Eqn.2) \\
&\dot{\rho_{\alpha\alpha}}=&\gamma_{c}(1+n_{c})(\rho_{11}\!+\!\rho_{22}+\rho_{33})\!-\!3\gamma_{c}n_{c}\rho_{\alpha\alpha}\!-\!\Gamma(1+\chi_{1})\rho_{\alpha\alpha},\\
&\dot{\rho_{\beta\beta}}=&\Gamma_{c}[N_{c}\rho_{bb}\!-\!(1+N_{c})\rho_{\beta\beta}]+\Gamma\rho_{\alpha\alpha},
\end{eqnarray*}
and
\begin{eqnarray*}
&\dot{\rho}_{a_{1}a_{1}}=&\gamma_{1h}[n_{1h}\rho_{bb}-(1+n_{1h})\rho_{a_{1}a_{1}}]+\gamma_{12}[n_{12}\rho_{a_{2}a_{2}}
                        \\&&-(1+n_{12}) \rho_{a_{1}a_{1}}+\gamma_{13}[n_{13}\rho_{a_{3}a_{3}}-(1+n_{13})\rho_{a_{1}a_{1}}] \\&&+\gamma_{1c}[n_{1c}\rho_{\alpha\alpha}-(1+n_{1c})\rho_{a_{1}a_{1}}],\\
&\dot{\rho}_{a_{2}a_{2}}=&\gamma_{12}[(1+n_{12})\rho_{a_{1}a_{1}}-n_{12}\rho_{a_{2}a_{2}}]+\gamma_{23}[n_{23}\rho_{a_{3}a_{3}}\\
                        &&-(1+n_{23})\rho_{a_{2}a_{2}}],~~~~~~~~~~~~~~~~~~~~~~~~~~~~~~~~~~~~~~~~~~~~~~~~~~~(Eqn.3)\\
&\dot{\rho}_{a_{3}a_{3}}=&\gamma_{3h}[n_{3h}\rho_{bb}-(1+n_{3h})\rho_{a_{3}a_{3}}]+\gamma_{23}[(1+n_{23})\rho_{a_{2}a_{2}}
                        \\&&-n_{23}\rho_{a_{3}a_{3}}]+\gamma_{13}[(1+n_{13})\rho_{a_{1}a_{1}}-n_{13}\rho_{a_{3}a_{3}}] \\&&+\gamma_{3c}[n_{3c}\rho_{\alpha\alpha}-(1+n_{3c})\rho_{a_{3}a_{3}}],\\
&\dot{\rho}_{\alpha\alpha}=&\gamma_{1c}[(1+n_{1c})\rho_{a_{1}a_{1}}-n_{1c}\rho_{\alpha\alpha}]+\gamma_{3c}[(1+n_{3c})\rho_{a_{3}a_{3}}
                        \\&&-n_{3c}\rho_{\alpha\alpha}]-\Gamma(1+\chi_{2})\rho_{\alpha\alpha},\\
&\dot{\rho}_{\beta\beta}=&\Gamma\rho_{\alpha\alpha}+\Gamma_{c}[\emph{N}_{c}\rho_{bb}-(1+\emph{N}_{c})\rho_{\beta\beta}],
\end{eqnarray*}
\noindent where \(n_{h}\)=\(\frac{1}{exp(\frac{E-E_{b}}{K_{B}T_{s}})-1}\), \(n_{ih(i=1,3)}\)=\(\frac{1}{exp(\frac{E_{a_{i}}-E_{b}}{K_{B}T_{s}})-1}\) describe the average numbers of photon with frequencies matching the transition energies from the VB state \(|b\rangle\) to the CB states \(|i\rangle_{(i=1,2,3)}\) in Fig.1(a), and  \(|a_{i}\rangle_{(i=1,3)}\) in Fig.1(b) at the temperature \(T_{s}\)=(300 +\(\Delta\))K, where \(\Delta\) stands for the temperature difference. \(n_{c}\)=\(\frac{1}{exp(\frac{E-E_{\alpha}}{K_{B}T_{s}})-1}\) and  \(n_{ic(i=1,3)}\)=\(\frac{1}{exp(\frac{E_{a_{i}}-E_{\alpha}}{K_{B}T_{s}})-1}\) are the thermal occupation numbers of ambient phonons at temperature \(T_{s}\).  \(N_{c}\)=\(\frac{1}{exp(\frac{E_{\beta}-E_{b}}{k_{B}T_{s}})-1}\) is the corresponding thermal occupation number at the ambient temperature \(T_{s}\) with the energy (\(E_{\beta}\)-\(E_{b}\)). \(n_{12}\), \(n_{13}\), and \(n_{23}\) represent the corresponding thermal occupations at the ambient temperature \(T_{s}\) with energy gaps (\(E_{a_{1}}\)-\(E_{a_{2}}\)), (\(E_{a_{1}}\)-\(E_{a_{3}}\)), and (\(E_{a_{2}}\)-\(E_{a_{3}}\)), respectively. The rates in Eqn. (2) and Eqn. (3) lead to a Boltzmann distribution for the level population \(|\alpha\rangle\) ( \(p_{\alpha}=exp(-\frac{E_{\alpha}-\mu_{\alpha}}{k_{B}T_{s}})\), \(\mu_{\alpha}\) is defined as the chemical potential of lead \(\alpha\)) when the thermal averages for the photon and phonon reservoirs are in a common temperature. We consider the initial condition to be a fully occupied ground state\cite{22}, i.e., \(\rho_{bb}\) = 1.

\section{Summary and Discussion}

In what follows, we calculate the steady solutions of Eqn.(2) and Eqn.(3) for the RRRs, \(\chi_{1}\) and \(\chi_{2}\).  Considering the cumbersome expression for  \(\chi_{1}\) and \(\chi_{2}\), we follow the numerical approach to carry out the discussion. And we use the following parameters, \(E_{\alpha}-\mu_{\alpha}\)=0.10 ev, \(E-E_{\alpha}\)=\(E_{\beta}-E_{b}\)=0.20 ev, \(\gamma_{c}\)=6 Mev, \(\Gamma\)=0.40 ev, \(\Gamma_{c}\)=0.15 ev\cite{21,22}. And other parameters are \(\gamma_{h}\)=0.62 ev in Fig.1(a), and \(\gamma_{1h}\)=0.62 ev ,\(\gamma_{3h}\)=0.45 ev, \(\gamma_{1c}\)=\(\gamma_{3c}\) =0.15\(\ast\)\((\frac{3}{2}\) +\(\sqrt{2}\)) ev, J=0.10 ev, \(\gamma_{12}\)=\(\gamma_{23}\) =\(\frac{1}{2}\)\(\gamma_{13}\)=0.15\(\ast\)\(\sqrt{2}\) ev in Fig.1(b).

Fig.2 plots the RRR (\(\chi_{i}\)) as a function of the temperature difference \(\Delta\)(K) with the energy gaps (\(E-E_{b}\)) being the control-parameters.
In Fig.2 (a), the linear curves decline with the temperature difference \(\Delta\)(K), and at room temperature (\(\Delta\)=0) the RRRs decrease with the increasing energy gaps (\(E-E_{b}\)). It indicates that in the quantum photocell with three uncoupled electron donors, the increasing circumstance temperature and energy gaps (\(E-E_{b}\)) can block the radiative recombination from the donors to acceptor. Therefore, the transported electrons to VB via the radiative recombination become less and more excited electrons were transported to perform useful work. These features suggest that the quantum photocell with three uncoupled electron donors operates at a slight higher temperature and wider band gap will bring out more excited electrons to perform useful work. However, we must take account of the fact that, increasing the band gap and environmental temperature will cause less absorption and much more phonon-electron scattering. Therefore, Fig.1(a) may indicate that the room temperature with a proper energy gap (\(E-E_{b}\)) is the optimal operating condition for the photocell with three uncoupled electron donors in the real environment.

\begin{figure}
\centerline{\includegraphics[width=0.45\columnwidth]{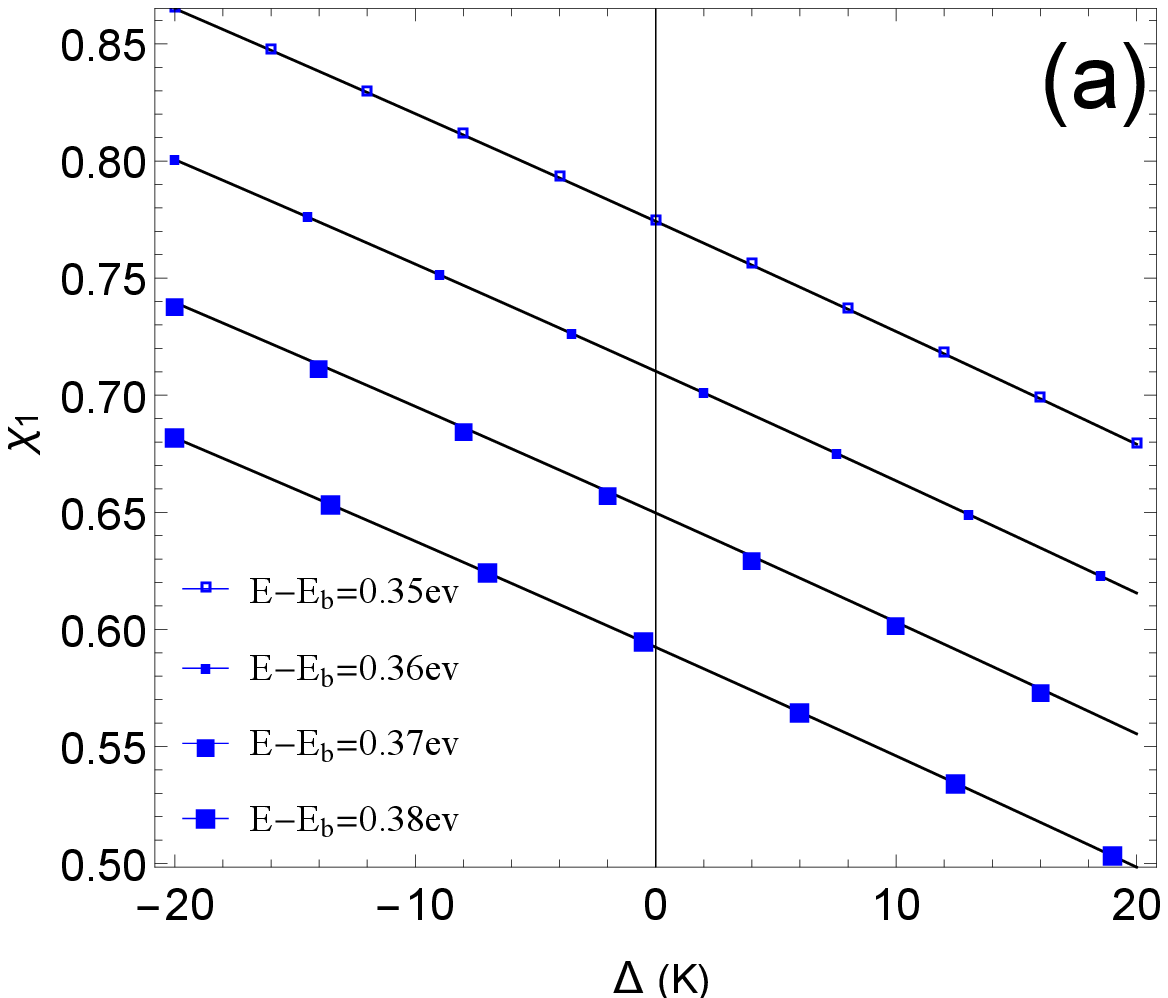}\includegraphics[width=0.46\columnwidth]{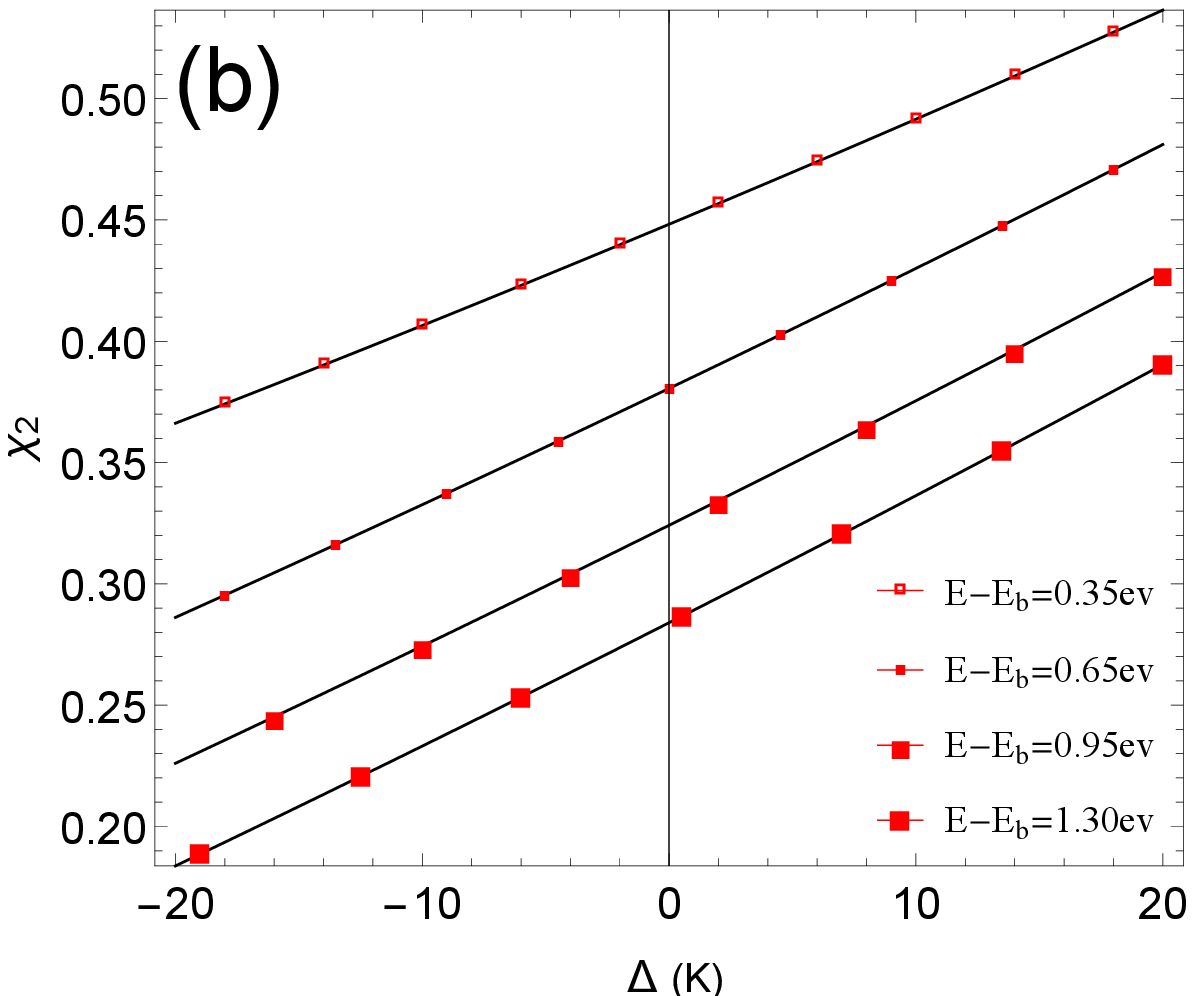}}
\caption{(Color online) (a) The RRR \(\chi_{1}\) with three uncoupled donors, (b) The RRR \(\chi_{2}\) with three dipole-dipole coupled donors as a function of the temperature difference \(\Delta\)(K) with different gaps (\(E-E_{b}\)) within three donors, respectively. \label{2}}
\end{figure}

However, a contrary result appears in the case of three dipole-dipole coupled donors in Fig.2(b). It notes that the increasing circumstance temperature promotes the RRRs but the energy gaps (\(E-E_{b}\)) can inhabit the RRR. And at the room temperature (\(\Delta\)=0) in Fig.2(b), the values of RRR are much smaller than those in Fig.2(a). The reason comes from their difference, i.e., the quantum photocell with three uncoupled donors in Fig.2(a) while with three dipole-dipole coupled donors in Fig.2(b). Therefore, in the condition of higher circumstance temperature, the quantum coherence generated from different electronic transport channels was strengthened in Fig.2(b), which holds back more electrons to perform work but radiate to the VB state \(|b\rangle\) directly in the donor molecules. Hence, just taking three dipole-dipole coupled donors into account, the increasing circumstance temperature inhabits the photon-to-charge efficiency. Although, Fig.2(b) demonstrates that the p-n junction with a slightly large energy gap(\(E-E_{b}\))
will decrease the possibility of the electronic radiative recombination in the photocell, increasing the band gap leading to less absorption should be considered in the photocell design.

\begin{figure}
\centerline{\includegraphics[width=0.44\columnwidth]{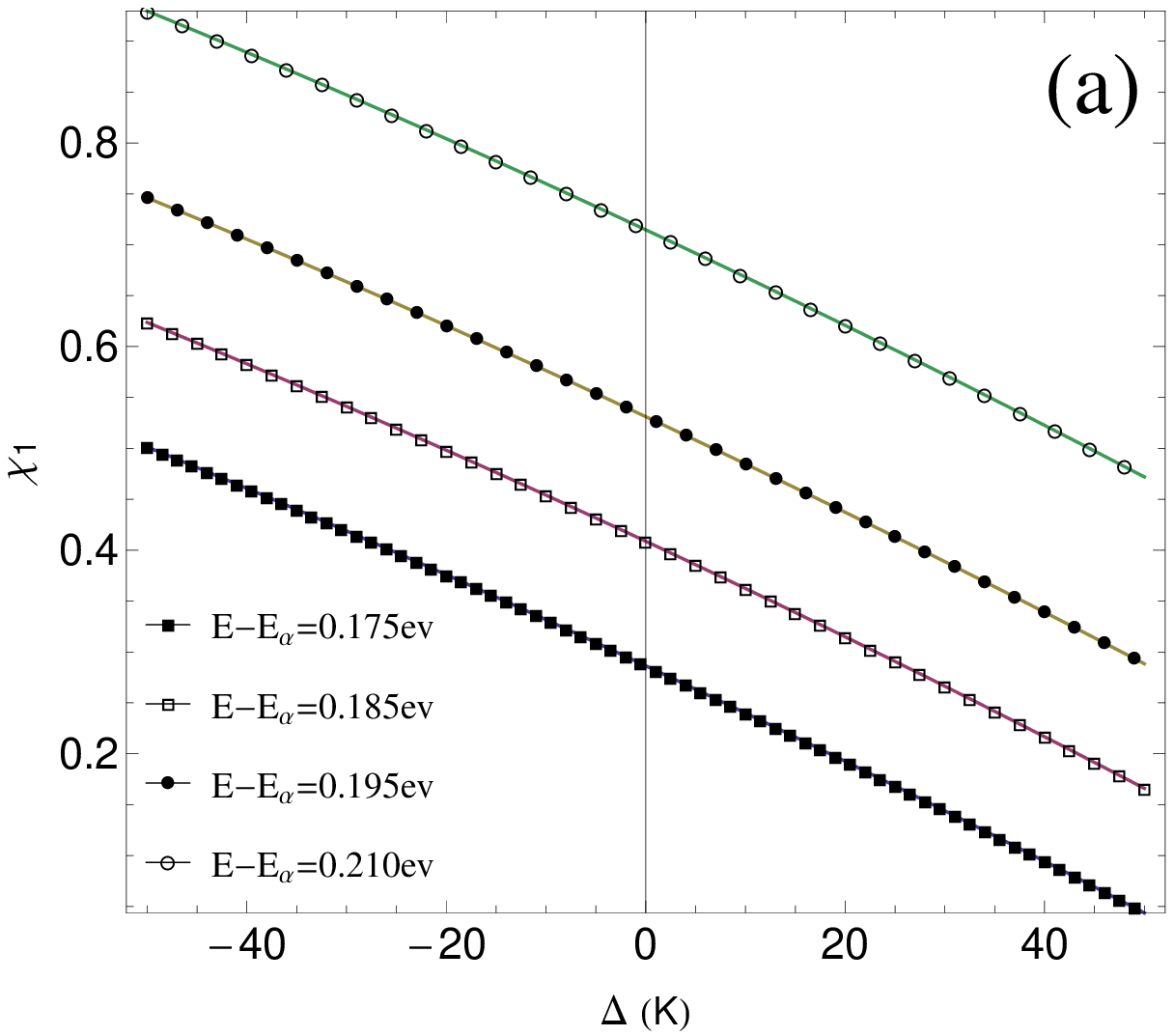}\includegraphics[width=0.45\columnwidth]{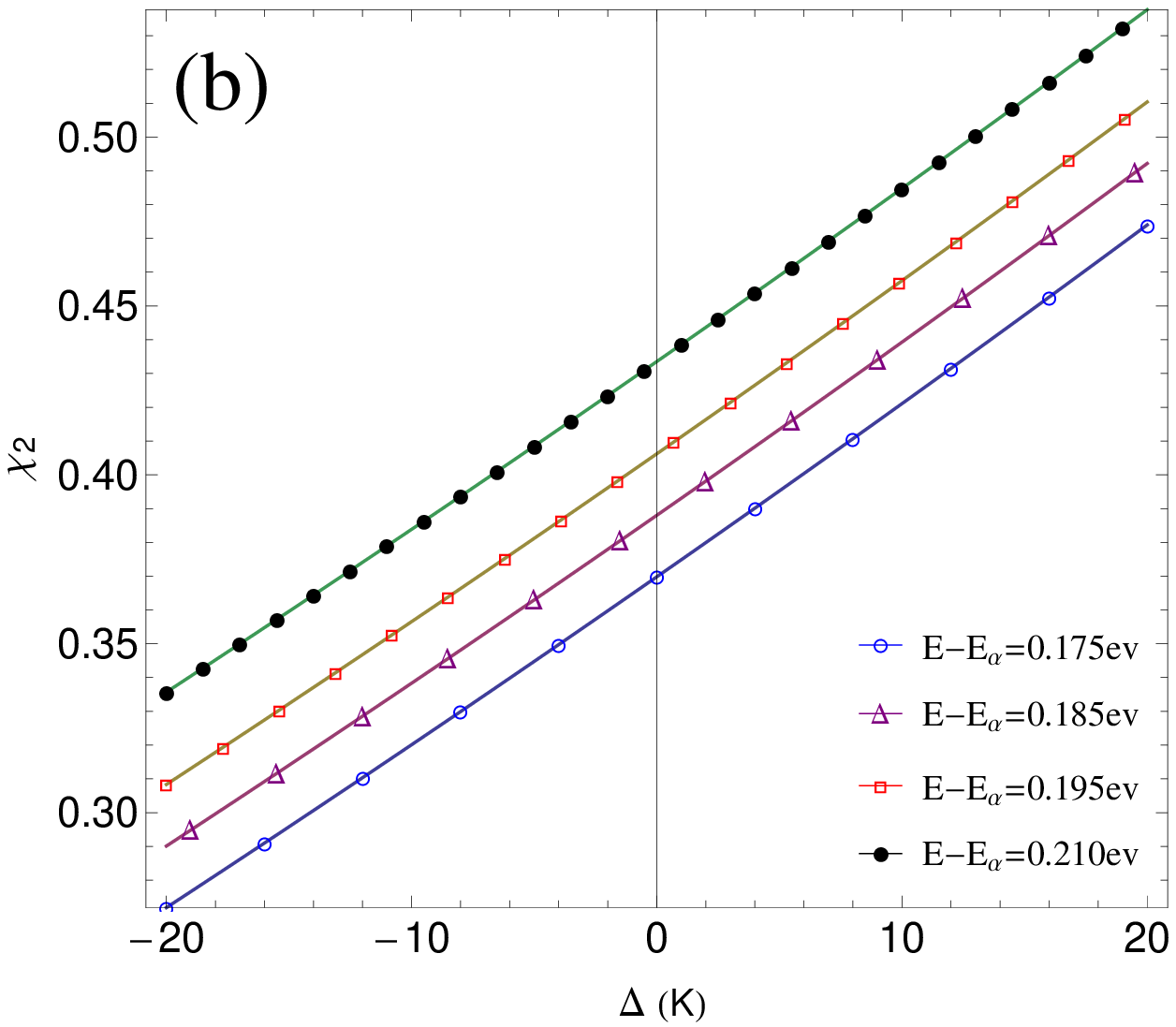}}
\caption{(Color online) (a) The RRR \(\chi_{1}\) with three uncoupled donors, (b) The RRR \(\chi_{2}\) with three dipole-dipole coupled donors as a function of the temperature difference \(\Delta\)(K)  with different energy gaps (\(E-E_{\alpha}\)) between the donors and the acceptor molecular. \label{3}}
\end{figure}

In the photocell, the gap between the donors and acceptor molecular, (\(E-E_{\alpha}\)) may be an interesting parameter to the RRR. Next, we carry out its influence on the RRR with the gaps (\(E-E_{b}\))=0.38 ev, and 0.7 ev in Fig.3(a) and (b), respectively, and other parameters are the same to those in Fig.2. The same physical features are shown in
Fig.3 as those in Fig.2. It notes that the ambient circumstance temperatures can suppress the RRR in Fig.3(a) with three uncoupled electron donors but enhance the RRR in Fig.3(b) with three dipole-dipole coupled electron donors, and at the room temperature (\(\Delta\)=0) the RRRs in Fig.3 (a) and (b) are smaller than those in Fig.2 (a) and (b), respectively. Not only that, but the RRR is inhabited by the decreasing gaps (\(E-E_{\alpha}\)) between the donors and the acceptor molecular in Fig.3. It is intuitively believed that it is not easy for the excited electrons transporting to the external load with a larger gap (\(E-E_{\alpha}\)) but radiative to the VB state  \(|b\rangle\). And the curves in Fig.3 both (a) and (b) demonstrate this conclusion.

\begin{figure}
\centerline{\includegraphics[width=0.45\columnwidth]{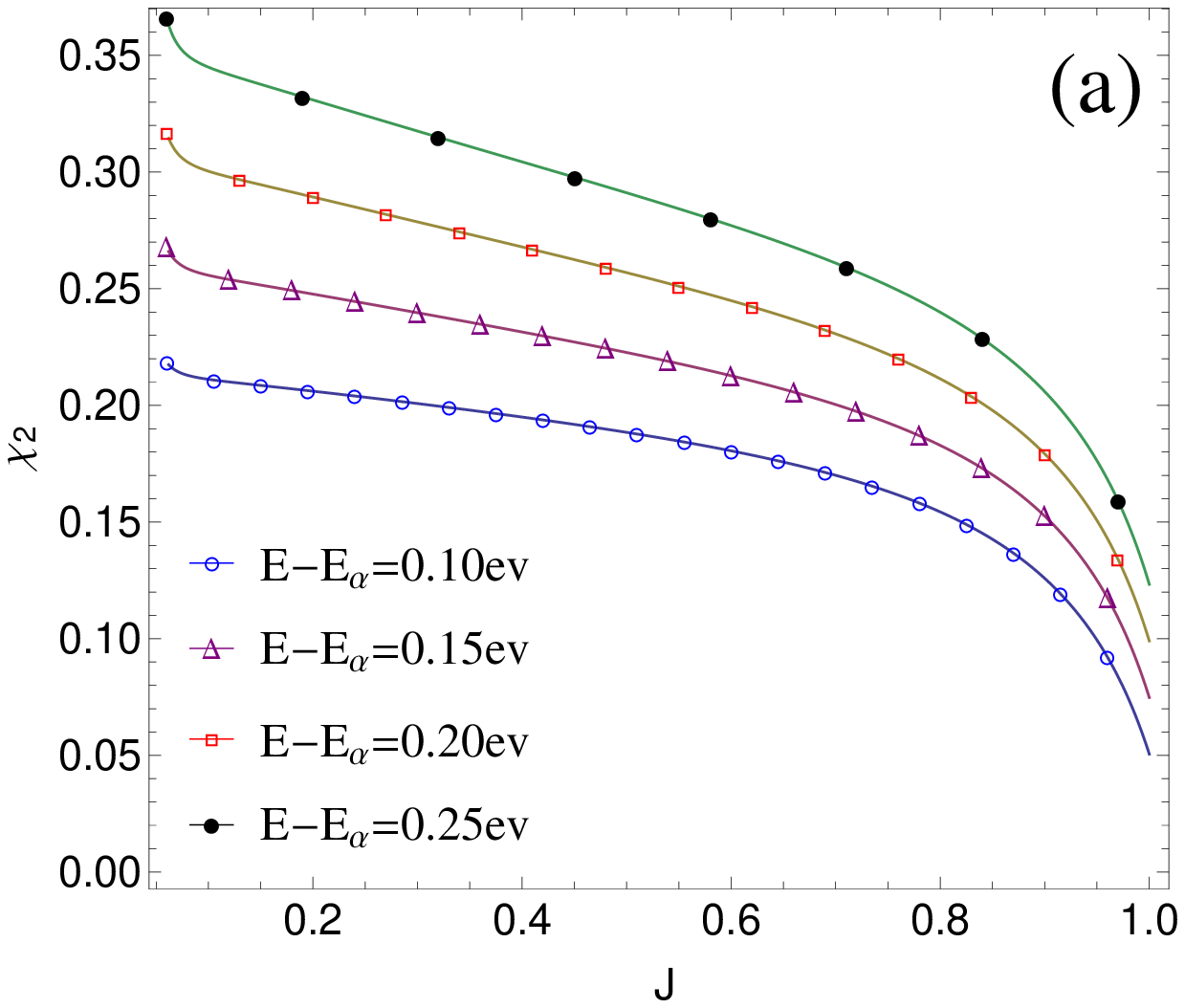}\includegraphics[width=0.44\columnwidth]{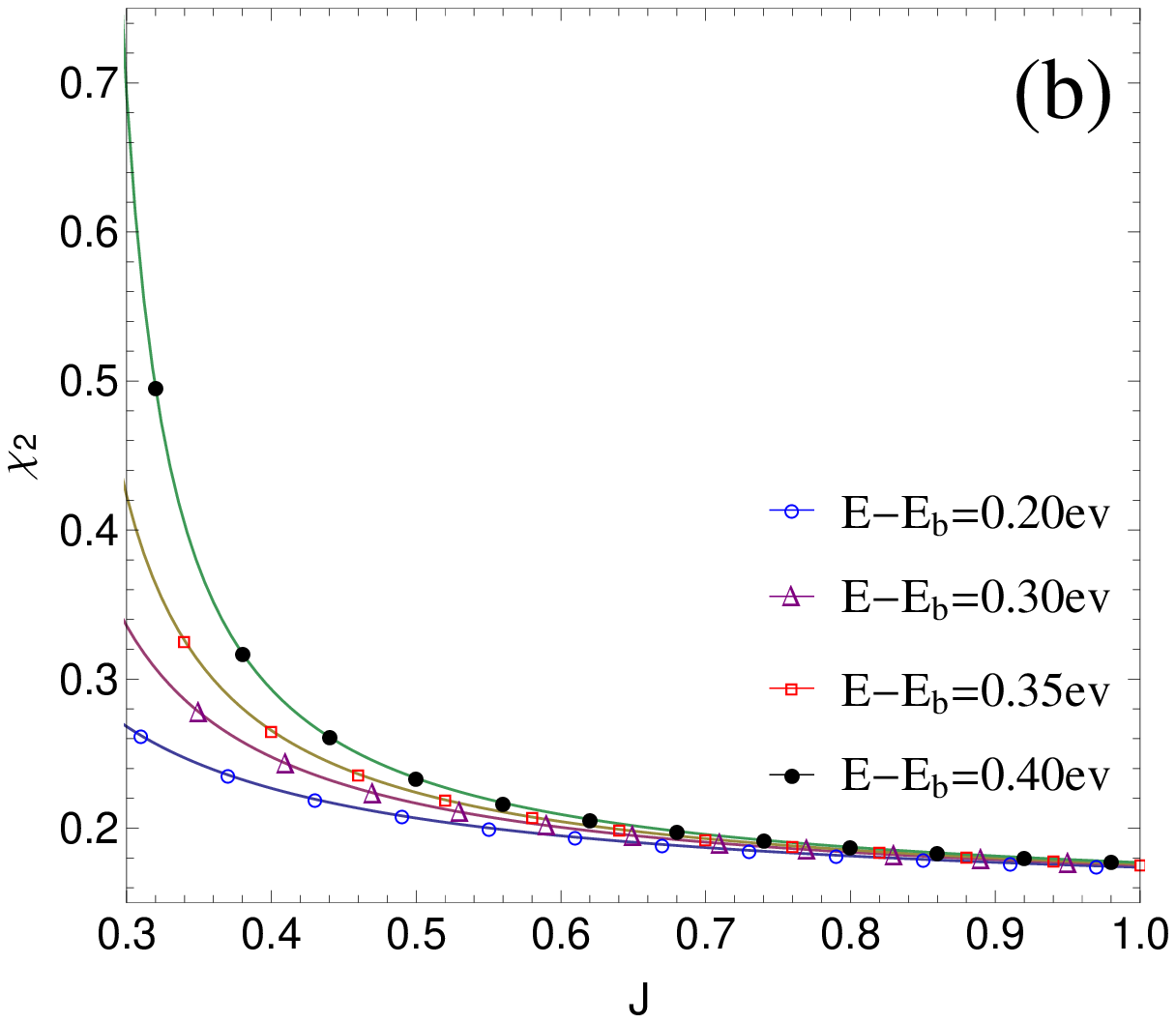}}
\caption{(Color online) The RRR \(\chi_{2}\) with three dipole-dipole coupled donors as a function of the electrostatic dipole-dipole coupling strength \(J\) with different energy gaps at room temperature.(a)(\(E-E_{b}\))=1.6 ev, (b) (\(E-E_{\alpha}\))=0.05 ev, and other parameters are the same to those in Fig.2. \label{4}}
\end{figure}

The increasing electrostatic dipole-dipole coupling strength \(J\) can bring out the stepped-up quantum interference between the different donors. To deeply investigate the quantum interference between the three dipole-dipole coupled donors in this quantum photocell, the following discussion about the RRR \(\chi_{2}\) versus the electrostatic dipole-dipole coupling strength \(J\) will be carried out in Fig.4 at room temperature. It shows that \(\chi_{2}\) monotonically decreases with the increasing \(J\) in both Fig.4(a) and (b), and the less gap (\(E-E_{\alpha}\)) in Fig.4(a) and (\(E-E_{b}\)) in Fig.4(b) generates the less RRR \(\chi_{2}\). However, the sharp decrease of \(\chi_{2}\) appears about in the range 0.8\(<\) \(J\) \( <\)1 in Fig.4(a) but appears about in the range 0.3\(<\) \(J\) \( <\)0.5 in Fig.4(b), and the final values of \(\chi_{2}\) are in close proximity to 0.05 in Fig.4(a), while the RRRs values infinitesimally approach to 0.2 in Fig.4(b). These results indicate that the increasing electrostatic dipole-dipole coupling strength \(J\) and the less gaps (\(E-E_{\alpha}\)) and (\(E-E_{b}\)) enhance the quantum interference between the three dipole-dipole coupled donors, which can inhabit the RRR \(\chi_{2}\) ultimately. And the RRR can be suppressed to a minimum but can't be canceled out\cite{17} due to the radiative upward and downward transition coexisting in the quantum photocell. When the gap (\(E-E_{b}\)) was fixed, the less gaps (\(E-E_{\alpha}\)) and the increasing electrostatic dipole-dipole coupling strength \(J\) enhance the quantum interference simultaneously, which leads to the suppressed \(\chi_{2}\). Then, the larger \(J\) generates the sharp decrease of \(\chi_{2}\) about in the range of 0.8\(<\) \(J\) \( <\) 1 in Fig.4(a). But in the case of (\(E-E_{\alpha}\)=0.05 ev), the increasing gap (\(E-E_{b}\)) cause less absorption solar photons. Therefore, the sharp decrease of \(\chi_{2}\) appears about in a smaller range about 0.3\(<\) \(J\) \( <\) 0.5 in Fig.4(b).

\section{ Possible experimental realization }

Up to now, we have investigated the features of RRR dependent the circumstance temperature, gaps and the electrostatic coupling strength of the dipole-dipole coupling \(J\) in this quantum photocell with three electron donors. Now let us suggest some potential experimental researches about this quantum photocell. First of all, two type of energy gaps discussed here display some significant results about the radiative recombination. It manifests that the semiconductor materials with appropriate energy gaps can effectively inhibit radiative recombination. Therefore, seeking a semiconductor material with suitable band gap in experiment may be an interested direction to suppress the radiative recombination. Secondly, the energy gap between the donors and acceptor may be another experimental investigation according our results, and best-effort to reduce this gap is a possible experimental realization for the suppressed RRR. How to align the donors of photocells for an intense electrostatic dipole-dipole coupling strength \(J\) in the manufacturing process may be another research field. The scenario proposed here may be a different approach for the efficient photon-to-charge conversion and deserve further experimental investigation.

\section{Conclusion}

To summarize, in this work we explored the RRR dependent ambient circumstance temperature, gaps and the electrostatic coupling strength of the dipole-dipole coupling \(J\) in a quantum photocell system with three electron donors. It showed that the RRR could be suppressed in this photocell with three uncoupled electron donors but be enhanced with three dipole-dipole coupled electron donors by the ambient circumstance temperatures, and the increasing energy gap in the donors, the decreasing gap between the donors and acceptor inhabited the RRR with three dipole-dipole both coupled and uncoupled electron donors. When the photocell was manipulated by the electrostatic dipole-dipole coupling strength \(J\) at room temperature, the RRR was suppressed to a smaller minimum by the gap between the donors and acceptor than those by different gaps in the donors. These results suggest an encouraging research direction to the high photon-to-charge conversion efficiency via the suppressed RRR in this quantum photocell, and some strategies for the suppressed RRR, such as seeking a semiconductor material with suitable band gap, minimizing the gap between the donors and acceptor and aligning properly the donors for an intense electrostatic dipole-dipole coupling strength \(J\) deserve the further experimental confirmation.

\begin{acknowledgments}
We thank the financial supports from the National Natural Science Foundation of China ( Grant Nos. 61205205 and 61565008 ), and
the General Program of Yunnan Applied Basic Research Project, China ( Grant No. 2016FB009 ).
\end{acknowledgments}





\bibliographystyle{99}

\begin{thebibliography}{00}
\bibitem{1} P. W$\ddot{u}$rfel, {\it Physics of Solar Cells}, (Wiley-VCH, Berlin), (2009).
\bibitem{2} W. Shockley and H. J. Queisser, {\it J. Appl. Phys.} {\bf 32}, 510 (1961).
\bibitem{3} M. A. Green, K. Emery, Y. Hishikawa, W. Warta, E. D. Dunlop, {\it Solar cell efficiency tables (version 48).Prog. Photovolt: Res. Appl.} {\bf 24}, 905, (2016).
\bibitem{4} U. W$\ddot{u}$rfel, M. Thorwart and E. R. Weber, {\it Quantum Efficiency in Complex Systems, Part II: From Molecular Aggregates to
            Organic Solar Cells, in Semiconductors and Semimetals},(vol. 85, Academic Press, San Diego), (2011).
\bibitem{5} O. D. Miller, E. Yablonovitch and S. R. Kurtz, {\it IEEE J. Photovoltaics}, {\bf 2 }, 303 (2012).
\bibitem{6} F. H. Alharbi, {\it J. Phys. D: Appl. Phys.}, {\bf 46}, 125102 (2013).
\bibitem{7} F. H. Alharbi and S. Kais, {\it Renewable Sustainable Energy Rev.}, {\bf 43}, 1073 (2015).
\bibitem{8} M. Daryani, A. Rostami, G. Darvish, M. K. Morravej Farshi, {\it Opt Quant Electron}, {\bf 49}, 255 (2017).
\bibitem{9} C. H. Henry, {\it J. Appl. Phys.}, {\bf 51 }, 4494 (1980).
\bibitem{10} A. Luque, P. G. Linares, E. Antolin, E. Canovas, C. D. Farmer, C.R. Stanley and A. Marti, {\it  Appl. Phys. Lett.}, {\bf 96}, 013501 (2010).
\bibitem{11} T. Nozawa and Y. Arakawa, {\it Appl. Phys. Lett.}, {\bf 98}, 171108 (2011).
\bibitem{12} A. Luque and A. Marti, {\it Phys. Rev. Lett.}, {\bf 78}, 5014 (1997).
\bibitem{13} J. Li, M. Chong, J. Zhu, Y. Li, J. Xu, P. Wang, Z. Shang, Z. Yang, R. Zhu and X. Cao, {\it Appl. Phys. Lett.}, {\bf 60}, 2240 (1992).
\bibitem{14} J. Bruns, W. Seifert, P. Wawer, H. Winnicke, D. Braunig and H. G. Wagemann, {\it Appl. Phys. Lett.}, {\bf 64}, 2700 (1994).
\bibitem{15} H. Kasai, H. Matsumura, {\it Sol. Energy Mater. Sol. Cells}, {\bf 48}, 93 (1997).
\bibitem{16} A. Luque, {\it J. Appl. Phys.}, {\bf 11}, 031301 (2011).
\bibitem{17} K. E. Dorfman, M. B. Kim, A. A. Svidzinsky, {\it Phys. Rev. A},{\bf 84}, 053829 (2011)
\bibitem{18} M. O. Scully, {\it Phys. Rev. Lett.}, {\bf 104}, 207701 (2010).
\bibitem{19} A. A. Svidzinsky, K. E. Dorfman and M. O. Scully, {\it Phys. Rev. A}, {\bf 84}, 053818 (2011).
\bibitem{20} M. O. Scully, K. R. Chapin, K. E. Dorfman, M. B. Kim and A. Svidzinsky, {\it Proc. Natl. Acad. Sci.}, {\bf 108}, 15097 (2011).
\bibitem{21} K. E. Dorfman, D. V. Voronine, S. Mukamel and M. O. Scully, {\it Proc. Natl. Acad. Sci.}, {\bf 110}, 2746 (2013).
\bibitem{22} C. Creatore, M. A. Parker, S. Emmott and A. W. Chin, {\it Phys. Rev. Lett.}, {\bf 111}, 253601 (2013).
\bibitem{23} K. E. Dorfman, K. E. Dorfman and M. O. Scully, {\it Coherent Optic Phenom},{\bf 1}, 42 (2013).
\bibitem{24} R. R. King, D. C. Law, K. M. Edmondson, C. M. Fetzer, et al., {\it Appl. Phys. Lett.}, {\bf 90}, 183516 (2007).
\bibitem{25} R. R. King, D. C. Law, K. M. Edmondson, et al., {\it Advances in Opto-Electronics},{\bf 2007}, 29523 (2007).
\end{thebibliography}

\end{document}